\documentclass[a4paper,12pt]{article}
\usepackage{color}
\usepackage[authoryear]{natbib}
\usepackage{graphicx}

\newcommand{\bx}{{\bf x}}
\newcommand{\bX}{{\bf X}}
\newcommand{\bw}{{\bf w}}

\newcommand{\bS}{{\bf S}}
\newcommand{\bd}{{\bf d}}

\newcommand{\bm}{{\bf m}}
\newcommand{\bzero}{{\bf 0}}
\newcommand{\bPsi}{\mbox{\boldmath $\Psi$}}

\newcommand{\bbeta}{\mbox{\boldmath $\beta$}}
\newcommand{\btheta}{\mbox{\boldmath $\theta$}}
\newcommand{\bmu}{\mbox{\boldmath $\mu$}}
\newcommand{\bSigma}{\mbox{\boldmath $\Sigma$}}
\begin{document}

\begin{center}
{\Large \textbf{A Bayesian semiparametric model for semicontinuous data}}

\vspace{1cm}

{\large Emanuela Dreassi\footnote{\textit{Corresponding author}: Dipartimento di Statistica, Informatica, Applicazioni ``G.
Parenti'' (DiSIA), Universit\`a degli Studi di Firenze, Viale
Morgagni, 59 - I 50134 Florence, Italy, dreassi@disia.unifi.it} and Emilia Rocco}\\

 \vspace{1cm}

{\small Dipartimento di Statistica, Informatica, Applicazioni ``G.
Parenti'' (DiSIA) Universit\`a degli Studi di Firenze}

\end{center}

\begin{abstract}
\noindent When the target variable exhibits a semicontinuous
behaviour (i.e. a point mass in a single value and a continuous
distribution elsewhere) parametric `two-part regression models' have
been extensively used and investigated. In this paper, a
semiparametric Bayesian two-part regression model for dealing with
such variables is proposed. The model allows a semiparametric
expression for the two part of the model by using Dirichlet processes. A
motivating example (in the `small area estimation' framework) based on
pseudo-real data on grapewine production in Tuscany, is used to
evaluate the capabilities of the model. Results show a satisfactory
performance of the suggested approach to model and predict
semicontinuous data when parametric assumptions (distributional and/or
relationship) are not reasonable.
\end{abstract}


\section{Introduction}
\label{introduction}

In many field of applications, in particular in biomedical and
economic studies, researchers encounter data that are either
continuous on the positive line or zero. In literature, such data,
in which the zeros are actual response outcomes and not proxies for
negative or missing responses, are referred as semicontinuous and
are usually handled by a non-standard two component mixture; whose
terms are a degenerate distribution (a point mass at zero) and some
standard distribution. This is realized by carrying out two
regression models, one for the mixing proportion (usually logit or
probit), the other for the mean of the standard distribution. The
latter is a conditional regression model that depends on the nature
of the data. Models of this type, are commonly called two-part
models and have large use in econometrics (\citealp{duan}, among
others) and for the analysis of longitudinal data in biomedical
applications (\citealp{olsen}; \citealp{berk}; \citealp{tooze};
\citealp{albert}; \citealp{gosh}). In this latter context, in order
to account for both the heterogeneity among individuals and the
possible correlation among subsequent observations on the same
individual, a cluster-specific random effect is usually enclosed in
both the parts of the model. More recently, the use of a two-part
random effects model has been suggested also in the context of small
area estimation (\citealp{pfeffermann}; \citealp{dreassi};
\citealp{chandra}).

In this paper, we propose a semiparametric Bayesian two-part model
where the model for the mixing proportion is semiparametric and the
model for the mean of the positive response outcomes is
nonparametric. The first one is a binary regression model in which
the commonly used parametric link function is replaced by a general
function for which a Dirichlet process (DP hereafter), centered
around a logistic distribution, is employed as prior distribution. In the second
one, a DP mixture of Normals is specified for the joint distribution
of the response and predictors. When model assumptions on
conditional distributions are not acclaimed, this approach, reducing
the need for parametric assumptions, brings down model specification errors.

Into a Bayesian paradigm, Bayesian nonparametric conditional
density regression models, have been widely used:
\cite{escobarfirst} and \cite{escobar} discuss computational issues
in DP mixture models where a parametric prior in a hierarchical
model is replaced by the nonparametric DP model; \cite{muller} and
\cite{dunson} face the problem of density regression using
Bayesian semiparametric and nonparametric approaches. Fitted
regression functions may be deduced as means of conditional
predictive distributions. The use of the DPs offers great
capabilities to furnish predictive distribution in a straight way;
so, if the main goal of the statistical analysis is prediction,
their use is very attractive. Another important characteristic of DP
mixtures is to define clusters on data. These latter seem very
appealing when structured data are the goal of the analysis.

Our suggestion is motivated by the fact that usually semicontinuous
data have a complex structure: they present a clustered (spatial
and/or temporal) structure; the positive values may have a highly
skewed distribution; the relation between the covariates and the
response may be not satisfactorily expressed by a model in which the
covariates enter the distribution of the response through a linear
function, and so on. In these cases, a parametric model cannot be
describe in appropriate manner the mechanism generating
the data and may be opportune relax parametric assumptions to
allow greater modeling flexibility. Bayesian nonparametric or
semiparametric models that allow achieving this flexibility are well
known in literature. However, it is unknown their use for the
analysis of semicontinuous data.

As motivating example we illustrate the application of suggested
semiparametric Bayesian two-part model to predict grapewine
production values on a small area estimation framework. Results
suggests that proposed semiparametric two-part model seems to be
able to: capture the particular relationship between response
variable and covariates included on the linear predictor,
discriminate between the two different mixture components, handle
the asymmetry of the data.

The paper is organized as follows. The suggested model is described
in Section \ref{model}. Section \ref{results} presents pseudo-real
data application. Final conclusion are reported in Section
\ref{conclusions}.

\section{The semiparametric Bayesian two-part model}
\label{model}

To account for semicontinuity of the response variable $Y$, it is
assumed that for each unit $i$ ($i=1, \ldots, n$) of the population
$Y_{i}=\delta_{i} \,\, Z_{i}$; $\delta_{i}$ is an indicator (i.e.,
it takes values 0 and 1 only) independent of the continuous random
variable $Z_{i}$. $\delta_{i}$ indicates that $Y_i$ came from the
continuous and not from the degenerate point-mass distribution.

We define a two part model. The first part predicts $\delta_{i} \mid
\bw_i$, the second part predicts $Y_i \mid \bx_i$; where $\bw_i$ and
$\bx_i$ represent two vector of explanatory variables.

In the first part, we consider a semiparametric Bernoulli regression
model for data $(\delta_i, \bw_i^\prime)$, where $\delta_i$ is a
binary response variable and $\bw_i$ an $r$-dimension vector of
predictors (intercept included). Parametric versions of this model
are characterized by the following assumption:
\[
P(\delta_i=1 \mid \bw_i, \btheta) = \mbox{E}(\delta_i=1 \mid \bw_i,
\btheta) = F_{\phi}\left[t(\bbeta^1, \bw_i)\right]
\]
where $F_{\phi}(\cdot)$ is a distribution function on the real numbers
(known up to a parameter $\phi$), called the inverse link
function in the context of generalized linear models, and $t(\cdot)$
is the index function, parameterized by $\bbeta^1$. Popular
parametric versions consider a linear index function $t(\bbeta^1,
\bw_i)=\bw_i^{\prime} \bbeta^1$, and a known cumulative distribution
function for $F_{\phi}$, thus allowing relatively simple treatment
of the finite regression parameters, $\btheta = \bbeta^1$. Following
\cite{jara}, we consider a latent variable representation $\delta_i
= I (V_i \leq \bw_i^\prime \bbeta^1)$ where $V_1, \ldots, V_n \sim
G^1$. We replace the parametric inverse link function $F_{\phi}$ by
a general distribution $G^1$ on which a DP prior is defined: $G^1
\sim DP(\alpha^1 G^1_0)$. We decided to center the prior around a logistic
distribution; i.e. the baseline prior distribution $G^1_0$ is a
Logistic$(V \mid 0, 1)$. To complete the model specification, a
Gamma$(a^1_0, b^1_0)$ for the precision parameter $\alpha^1$ (see
\citealp{escobar}) and a $\mbox{Normal}_r(\bbeta^1_0,
\bS_{\beta^1_0})$ for regression coefficients $\bbeta^1$ are given.
A Metropolis-Hastings step is used to sample the full conditional
distribution of the regression coefficients and precision (see
\citealp{jara}).

The second part is carried on just for positive values data $j =
1, \dots, m$, where $m < n$. A DP mixture of Normal distribution
(\citealp{escobar}) for the conditional density estimation on $Z_j
\mid \bx_j$ is used. According to \cite{muller} we specified a DP
mixture of Normals for the joint distribution of the response and
predictors and we looked at the induced conditional regression. Even
if, in the original paper, \cite{muller} focussed on the mean
regression function their method can be used to model the
conditional density of the response giving the predictors (see
\citealp{dunson}). Let $Z_j$ and $\bX_j$ be the response and the $p$
dimensional vector of continuous predictors, respectively. Further,
let $\bd_j = (z_j; \bx_j^\prime)^\prime$, with $j=1, \ldots, m$ and
$k=p+1$ dimension. The model for the joint distribution of the
response and predictors is: $\bd_j \sim \mbox{Normal}_k (\bmu_j,
\bSigma_j)$, with \textit{iid} distributions for $\left(\bmu_j,
\bSigma_j\right)$, $j=1, \ldots, m$. For each $j$, $\left(\bmu_j, \bSigma_j\right)
\sim G^2$ and $G^2 \sim \mbox{DP}(\alpha^2 G^2_0)$. The prior for
the baseline distribution $G^2_0$ is the conjugate Normal - Inverted
Wishart distribution
\[
G^2_0 \equiv \mbox{Normal}_k(\bm_{1}, k_0^{-1} \bSigma) \quad
\mbox{Inverse Wishart}_k (\nu_1, \bPsi_1)
\]
The model specification is completed when the independent priors are
given: $\alpha^2 \sim \mbox{Gamma}(a^2_0, b^2_0)$, $\bm_{1} \sim
\mbox{Normal}(\bm_{2}, \bS_2)$, $k_0 \sim \mbox{Gamma}(\tau_1/2,
\tau_2/2)$ and $\bPsi_1 \sim \mbox{Inverse Wishart}_k (\nu_2,
\bPsi_2)$.

This second part of the model defines a weight dependent mixture
models:
\[
f(z)= \sum_{l=1}^\infty \omega_l(\bx) \mbox{Normal}(\beta^2_{0l}+
\bx^\prime \bbeta^2_l, \sigma_l^2)
\]
where
\[
\omega_l(\bx)=\frac{\omega_l
\mbox{Normal}_p(\bmu_{2l},\bSigma_{22l})}{\sum_{q=1}^\infty \omega_q
\mbox{Normal}_p(\bmu_{2q},\bSigma_{22q})}
\]
with $\beta^2_{0l}=\mu_{1l}- \bSigma_{12l} \bSigma_{22l}^{-1}
\bmu_{2l}$, $\bbeta_l^2= \bSigma_{12l}\bSigma_{22l}^{-1}$ and
$\sigma_l^2=\sigma_{11l}^2 - \bSigma_{12l}\bSigma_{22l}^{-1}
\bSigma_{21l}$. The weights $\omega_l$ follow a DP stick-breaking
construction and the other components derive from the standard
partition of the vectors of means and variance and covariance
matrices given by $\bmu_l=(\mu_{1l}, \bmu_{2l})^\prime$ and
$\bSigma_l=\left( \begin{array}{cc}
                   \sigma_{11l}^2 & \bSigma_{12l} \\
                   \bSigma_{21l} & \bSigma_{22l}
                 \end{array}
\right).$

To complete the model, we assigned the values to hyperparameters. In
the following application, for example, we considered $a^1_0=2$,
$b^1_0=1$, $\bbeta^1_0=\bzero$, $\bS_{\beta^1_0}$=diag$_r(10000)$,
$a^2_0=10$, $b^2_0=1$, $\nu_1=\nu_2=4$, $\bm_2=(\bar{z},
\bar{x})\prime$, $\tau_1=6.01$, $\tau_2=3.01$ and
$\bS_2=\bPsi_2^{-1}$=0.5 $\bS$, where $\bS$ is the sample
variance-covariance matrix for the response and predictor.

To sum up, the two-part model parameters are the followings. The
precision parameters of the DPs, respectively $\alpha^1$ and
$\alpha^2$ for the first and the second part of the model, and the
number of clusters that the DPs induce. From the first part of the
model, the $\bbeta^1$ regression coefficients. From the second part
of the model, in the case that just one predictor is included (as on
the motivating example considered): the mean $\bm_1$ of the Normal
component of the baseline distribution $G^2_0$ as a bivariate vector
with elements $m_{1,z}$ and $m_{1,x}$; the scale matrix $\bPsi_1$ of
the inverted Wishart part of the baseline distribution $G^2_0$ as a
$2 \times 2$ symmetric matrix with elements $\psi_{1,z}$,
$\psi_{1,x}$ (on the diagonal) and $\psi_{1,zx}$ (out diagonal).
Finally, the scale parameter $k_0$ of the Normal part of the
baseline distribution $G^2_0$.

From the first part of the model, we obtained
$\mbox{E}\left(\delta_i=1 \mid \bw_i \right)$. From the second part
of the model, we obtained in straight way the predictive distribution
$f(z_i \mid \bx_i)$. Finally, in order to obtain the predictive distribution $f(y_i \mid \bx_i, \bw_i)$, according to the
parametric semicontinuous two-part models standard practice, we
considered the product $f(z_i \mid \bx_i) \, \,
\mbox{E}\left(\delta_i=1 \mid \bw_i \right)$.

\section{Motivating example: the pseudo-real data on grapewine production in Tuscany}
\label{results} In this section, we present an empirical evaluation
of the proposed modeling by analyzing some pseudo-real agricultural
data on grapewine production in Tuscany. A specific crop production
is a typical case of semicontinuous output variable and the
grapewine production in Tuscany does not contradict this assertion.

Our data come from two survey conducted by the Italian Statistical
Institute (ISTAT): the Fifth Agricultural Census performed in 2000
(hereafter census2000) and the Farm Structure Survey performed in
2003 (hereafter FSS2003). Both census2000 and FSS2003 are a
reiteration of two surveys routinely conducted by ISTAT, ten-yearly
and two-yearly respectively, in order to monitor trends and
transitions in the structure of farms, but also to model the impact
of external developments or policy proposals. For both the census
and the sample survey the unit of observation is the farm for which
surface areas (measured in hectares) allocated to different crops,
as well as many other socioeconomic variables, are recorded. Two
other important features of our data worthy of mention are the
following: in FSS2003, as in all reiteration of this survey up until
2005, the production of each crop (quantity in quintals) has been
observed and in census2000, for the first time, spatial information
was collected. It consists of the geographical coordinates of each
farm's administrative center.

Notwithstanding the FSS2003 uses a sample of more than 50,000 farms
on the whole Italy, it is conceived to provide reliable estimates at
regional level only. However, for each region it is often required
to produce estimates even at sub-regional level, at least for its
main crops. Direct estimates are not well-suited because only a
small sample is available. Hence, we must refer to indirect
estimators, which exploit the available variables collected at the
census2000 as auxiliary variables. In fact, the lag time between the
response and the auxiliary variables can be assumed to be
negligible, because of the high correlations, among the auxiliary
variables measured for the sampled farms in both the years 2000 and
2003. Nevertheless, it is obviously possible that some farms have
been changed their activity.

\begin{figure}
  \centering{\includegraphics[width=0.75\textwidth]{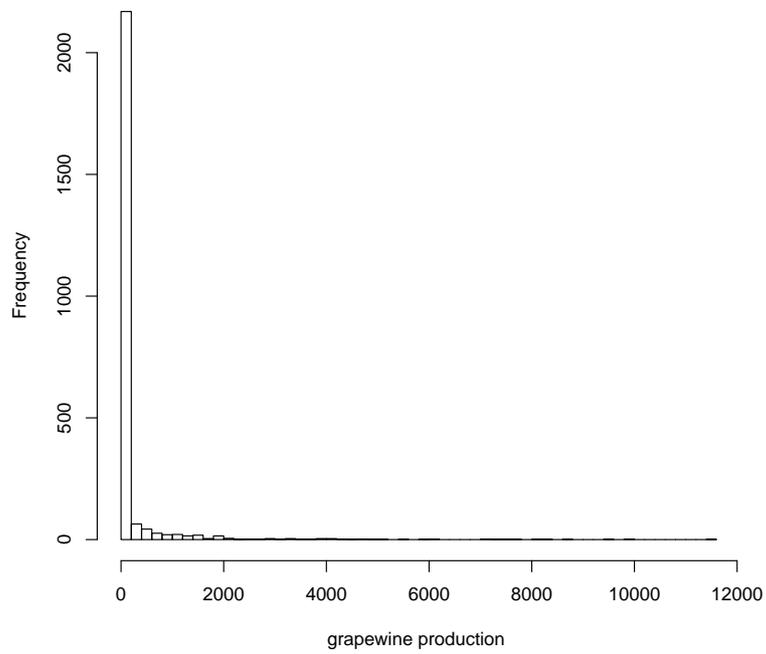}}
     \caption{Data on grapewine production: the histogram for the 2450 units from the FSS2003 survey.}
    \label{figure1}
\end{figure}

Here we consider only the FSS2003 part for the
Tuscany region; hence 2450 farms. A large number of these farms
(1489) do not produce grapewine, while only a few (961)
produce the majority of the total production in Tuscany and the
distribution of the positive grapewine production in these farms is
highly skewed. Figure \ref{figure1} shows the semicontinuous nature
of the grapewine production variable.

In this paper, we focused on the ability of the proposed model to
predict grapewine production using some auxiliary variables; this
for small area estimation end. In a small area estimation setting,
once we have obtained the predictive distribution, we can extract a
prediction for the out of sample units and finally obtain, using a
plug-in estimator (that combines predicted and observed values), the
area mean or total estimates (for different area levels).

In order to fit the model we decided to use only a part of the whole
dataset of 2450 units. Hence, we randomly split the FSS2003
sample into two parts; these contain respectively 816 and 1634 farms
(1/3 and 2/3 of data). We used information on the
816 farms to predict the grapewine production for the others 1634
farms (for which the production from FSS2003 survey is registered,
hence known). In this way comparison between the true values and the
predicted ones from the suggested model is feasible. In the
following we denoted as `observed' the grapewine production from
FSS2003 for the 816 farms, and `true' and `predicted' respectively
the grapewine registered from FSS2003 and predicted from the
suggested model for the others 1634 farms.

The selection of the covariates to be included in each of the two
parts of the model, among several socioeconomic variables available
at census2000, was first performed using an explorative analysis. We
conducted  a preliminary parametric analysis on the data. A Logistic
model has been first fitted to these data and the choice between
alternative models (including different covariates) have been made
basing on AIC (Akaike Information Criteria). For the first part of
the model, four auxiliary variables are considered: presence/absence
of surface allocated to grape wine, a relative measure of the latter
on the overall cultivated surface, to be or not a grapewine seller,
the slope of the farm's ground. Because in census2000 also the geographical coordinates have been registered,
we easily obtained this latter covariate by merging slope information on a grid of geographical coordinates.
We decided to include just one covariate
in the second part of the model for a simpler analysis of the
results. The `more explicative' covariate is the surface allocated
to grape wine.

For the estimation of the models, via MCMC simulation methods, we
used the \texttt{DPbinary} and \texttt{DPcdensity} functions from
the library \texttt{DPpackage} of the \texttt{R package} (see
\citealp{jara2}). A sensitivity analysis on the hyperparameters values choice has
been carried out. Convergence has been checked by
\citet{gelmanrubin} convergence diagnostic criterion. The algorithm
seems to converge after a few thousand iterations. However, given
the very high number of (non monitored) parameters in the model, we
decided to discard the first 200,000 iterations (burn-in) and to
store 2000 samples (one each 100) of the following 200,000
iterations.

From the first part of the model, we obtained $\mbox{E}(\delta_i=1
\mid \bw)$, hence the probability to have positive grapewine
production for each $i$. The sum, over the farms, is 332.6; this
suggest good performance of the first part of the model because 333
are the farms with positive grapewine production (483 with zero
production). Moreover, on the 816 units modelled, the mean for units
with zero production is 0.16, whereas for those with positive
production is 0.76. To evaluate predictive capabilities of the first
part of the model, we considered a cut-off of 0.5 over $\mbox{E}\left(
\delta_i = 1 \mid \bw \right)$; this to the end of classify (by prediction)
the farms with zero or positive production. On the whole 816 units,
the farms that have zero production and are rightly classified are the
51\% whereas those wrongly classified are the 8\%. Moreover, the farms
that have positive production and are correctly classify are the
38\%, but those uncorrected classified are the 3\%. To sum up,
according to a cut-off of 0.5, the 89\% seem to be rightly classified
and 11\% are misclassified.

\begin{figure}
  \centering
     \includegraphics[width=0.75\textwidth]{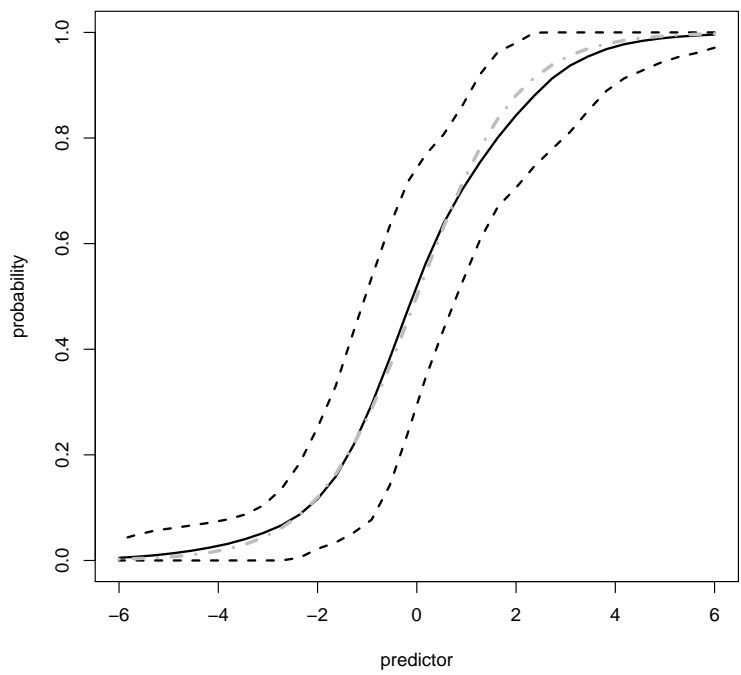}
   \caption{Data on grapewine production: estimated link function (black solid line) and its 95\% credibility intervals (black dashed lines)
   \textit{versus} parametric logistic link function (gray dashed line).}
    \label{figure2}
\end{figure}

Figure \ref{figure2} describes the nonparametric link function
estimated from the first part of the model. It seems to be less
steep for large values of the predictor $\bbeta \bw_i^\prime$
respect to the prior distribution (i.e. the logistic).

Table \ref{table1} reports some descriptive values of the estimated
posterior distribution for the parameters of the semiparametric
Bayesian two-part model.

\begin{figure}
  \centering
     \includegraphics[width=1\textwidth]{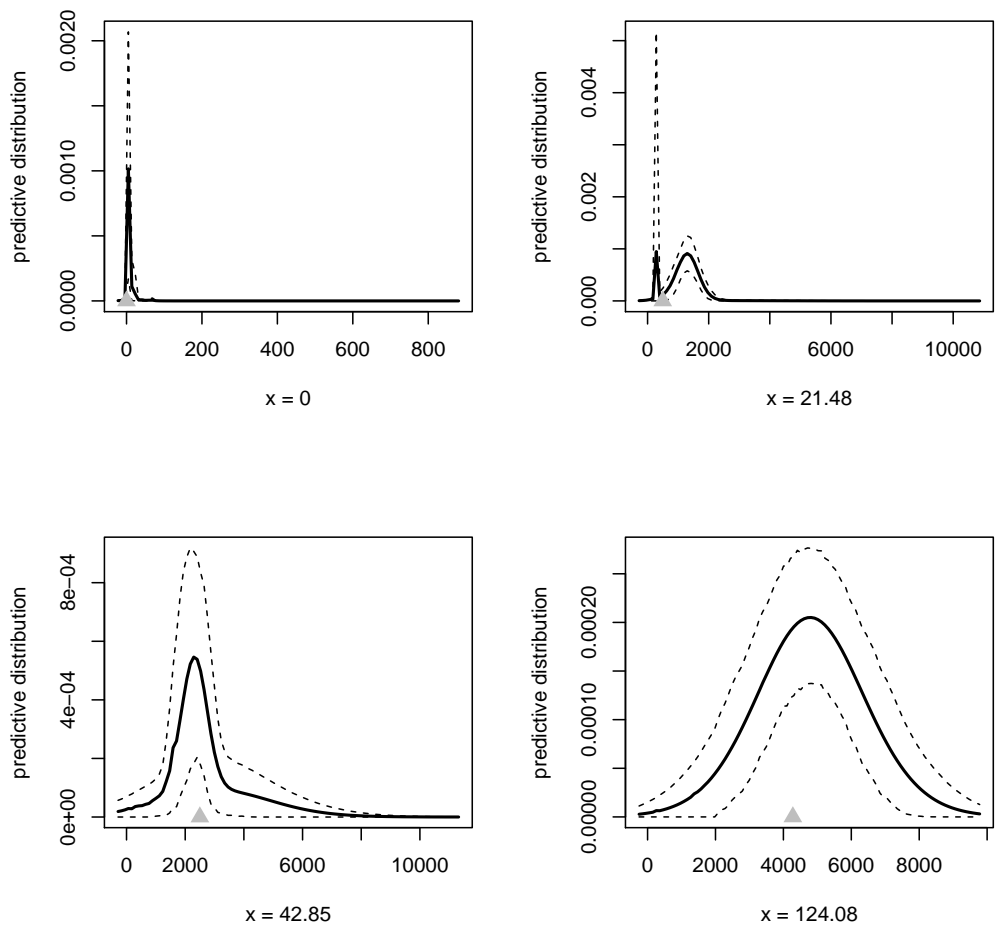}
   \caption{Data on grapewine production: conditional predictive distribution $f(y_i \mid x_i, \bw_i)$ (solid line)
   with its 95\% credibility intervals (dashed lines); true grapewine production value (gray triangle).}
    \label{figure3}
\end{figure}

\begin{figure}
  \centering
     \includegraphics[width=1\textwidth]{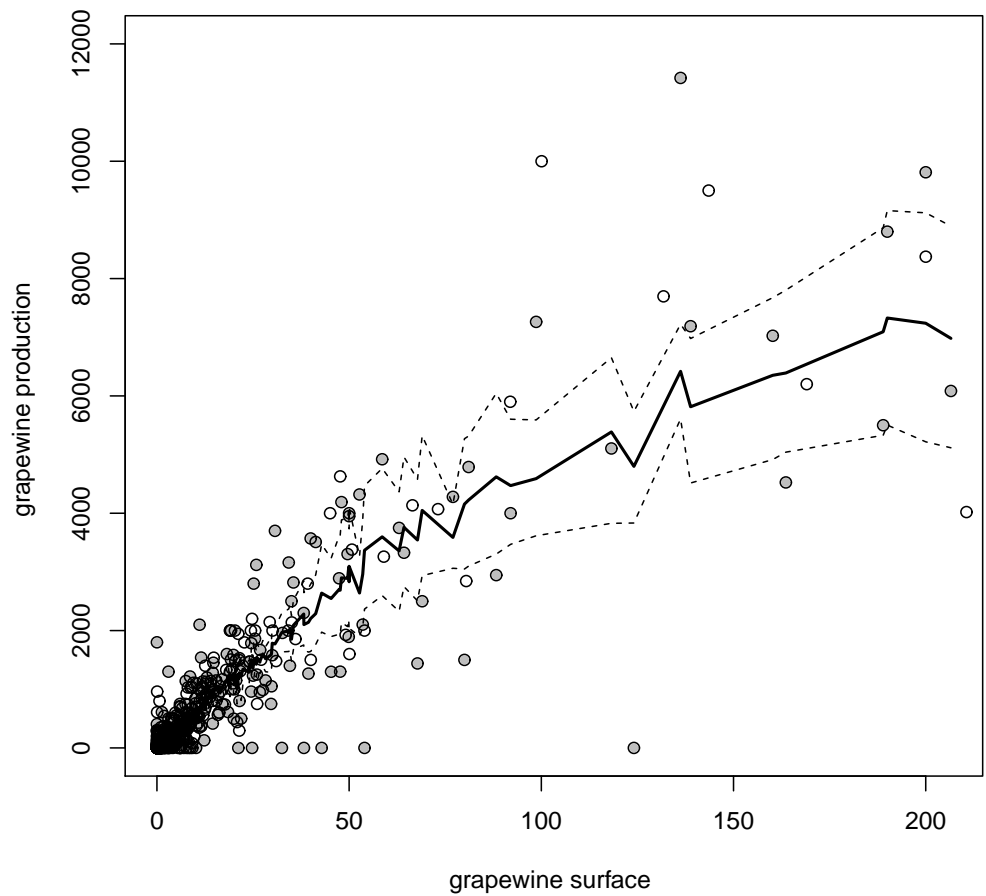}
   \caption{Data on grapewine production: (white circle) observed data; (gray circle) true value;
   (solid line) fitted regression prediction function with its 95\% credibility intervals (dashed lines).}
    \label{figure4}
\end{figure}

Figure \ref{figure3} shows estimated conditional predictive
distributions $f(y_i \mid x_i, \bw_i)$ for selected values of the
covariate $x_i$ (surface allocated to grapewine from census2000)
from the 1634 farms; we decided to have a description for units with
$x_i=0, 21.48, 42.85, 124.08$. The gray triangle represents
the true grapewine production value for the farm with
surface allocated to grapewine $x_i$. The proposed semiparametric two-part
model seems to have good prediction performances. Figure
\ref{figure4} describes the prediction for the 1634 farms: the fitted
regression function $\mbox{E}(y_i \mid x_i, \bw_i)$ obtained from
the suggested model. We noted a particular non-linear relation
between surface allocated to grapewine production from census2000
and grapewine production from FSS2003. Differences between
prediction and true values are high when not support from data is
given to estimate predictive distributions. Note that we
obtained the predictive distribution for the semicontinuous variable by multiply a
predictive distribution (obtained from the second part of the model)
and an expected value (obtained from the first part). As a consequence, the
95\% credibility intervals, defined for $f(y_i \mid x_i, \bw_i)$, take
into account only for the variability on estimates from the second
part of the model.

\section{Conclusions}
\label{conclusions} The results, obtained from the application to data on grapewine production,
show that the proposed methodology provides a reasonable and useful alternative to existing
methods when assumptions of the parametric model are not valid. Moreover, an
appealing feature of the suggested model is to work directly on the
conditional predictive distributions. Despite the fact that the
proposed methodology provides encouraging results, further research
is necessary. To start with, thanks to the clustering properties of the DP mixture, the inclusion of a correlation structure could be overcome;
but anyway we can provide the explicit inclusion of some (time or space) correlation structures. Moreover, regarding the  95\% credibility intervals,
since we have not take into account for the variability of estimates from the first part of the model, we can extend the model to cope with this variability.
Finally, we can argue that because we use a mixture model on the second part,
we can include on it also the degenerate component, so to consider a `single-part' hierarchical nonparametric Bayesian model.

\section*{Acknowledgement}
This work has been supported by project PRIN-2012F42NS8 ``Household wealth
and Youth unemployment: new survey methods to meet current challenges''.

\newpage

\begin{table}
\begin{center}
\caption{Data on grapewine production: estimate of the parameters of
the semiparametric Bayesian two-part model with their 95\%
credibility intervals (95\% CI).}\label{table1}
\begin{tabular}{lcc}
\hline
parameter &  mean   &  95\% CI\\
\hline
$\beta^1_0$ (intercept)&  -2.622   &-4.606 ; -1.260\\
$\beta^1_1$ (surface presence/absence) &  3.823   &  2.163 ; 5.818           \\
$\beta^1_2$ (relative grapewine surface) & 1.744    & 0.271 ;  4.019            \\
$\beta^1_3$ (to be seller yes/no) & 0.826   & 0.174 ; 1.853     \\
$\beta^1_4$ (slope) & -0.638   & -1.483 ; 0.025  \\
$\alpha^1$ &  16.525    &  7.057 ; 26.895          \\
number of clusters from DP first part& 79.194 & 41 ; 114\\
\hline
$m_{1,z}$&  168.400  & 66.740 ;  291.900  \\
$m_{1,x}$&  3.242  &  0.454 ;  6.584  \\
$k_0$ & 0.011  &  0.002  ; 0.032   \\
$\psi_{1,z}$& 0.002  &  0.001 ;  0.003  \\
$\psi_{1,x}$& 1.422  &  0.797 ;   2.424 \\
$\psi_{1,zx}$& -0.017     &   -0.039 ; -0.002  \\
$\alpha^2$& 6.917  & 4.017  ; 10.658\\
number of clusters from DP second part& 24.173 & 16 ; 33\\
\hline
\end{tabular}
\end{center}
\end{table}

\end{document}